\documentclass[fleqn,twoside]{article}
\usepackage[headings]{espcrc2}

\readRCS
$Id: espcrc2.tex,v 1.2 2004/02/24 11:22:11 spepping Exp $
\usepackage{graphicx}
\usepackage[figuresright]{rotating}

\newcommand{\AmS}{{\protect\the\textfont2
  A\kern-.1667em\lower.5ex\hbox{M}\kern-.125emS}}

\title{B-identification for Level 2: The Silicon Track Trigger at D\O\ }

\author{Sascha Caron \address[MCSD]{NIKHEF, 
        Amsterdam \\ 
        The Netherlands \\
	scaron@nikhef.nl}
\thanks{ I like to thank the organisers for a very enjoyable conference and
the STT and Level-2 groups  at D\O\ for all their help. I acknowledge the support of a 
Marie Curie Intra European Fellowship in the 6th EU Framework Programme.
}
for the D\O\ collaboration
}
               
\runtitle{The Silicon Track Trigger at D\O\ }
\runauthor{S. Caron}

\begin{document}

\begin{abstract}
This article describes the Silicon Track Trigger (STT) which has been fully
commissioned in 2004
at the D\O\ experiment. The STT allows to enrich
 already at the second trigger level 
the data sample with events containing B-mesons. The STT achieves this by
providing within about $50\mu s$ tracks
with an impact parameter resolution of around $50\mu m$.
The article shows preliminary results
of the trigger performance and presents a fast $b$-identification algorithm for
the second trigger level.

\vspace{1pc}
\end{abstract}

\maketitle

\section{Introduction}
The discovery of the Higgs boson is one of the main objectives 
of high energy physics today. 
Especially at the Tevatron this is a very difficult task, because
it requires 
to study all possible decay channels with the best achievable efficiency.
The overwhelming amount of light quarks produced by QCD processes 
swamps interesting signals with b-quarks like
$p\overline{p} \rightarrow H \rightarrow b \overline{b}$ 
or the important calibration process $p\overline{p} \rightarrow Z \rightarrow b\overline{b}$. Even
Higgs processes 
with the associated production of additional b-jets or neutrinos 
($HZ\rightarrow b\overline{b} \nu \overline{\nu}$)
have huge background from light quark jets
, such that the limited trigger bandwidth
of an experiment results in a loss of some of those events.
Physics topics as the study of low $P_T$ B-physics
face similar performance issues.

If the early trigger levels of an experiment
cannot discriminate between light quarks and b-quarks,
both will be reduced by an equal factor. 
The Silicon Track Trigger (STT)\cite{stt}  , however, can early
recognize events where a $b$-quark was produced and
dominantly pass those events, reducing the background effectively.
B mesons lead to
trajectories of the order of millimeter before they decay.
The trajectories of the decay products do not point back to
the vertex of the primary interaction. Hence the measurement of the
impact parameter 
allows the separation of interesting $b$-events from events containing
only lighter quarks. The impact parameter (or distance of 
closest approach) is the minimum distance between the primary interaction
point and the particles trajectory.
Note that such a method,  unlike the usual selection
of B-meson via muon decays, works for all 
decay modes and provides a less biased
sample of decay modes in the selected $b$-event sample.

In RunII the D\O\ experiment also selects $b$-events 
using the high resolution 
Silicon Microstrip Detector (SMT) to reconstruct the tracks of
the charged particles in the event and 
by feeding the track information into  
$b$-identification algorithms.
Without the Silicon Track Trigger 
this can only be performed in offline analyzes and 
the third trigger level.
Since early 2004 the STT provides the capability of a fast selection of 
events with large impact parameter. This allows B-meson identification 
already at the second trigger level.

\section{The STT in the D\O\ trigger framework}
The D\O\ experiment triggers events in three stages \cite{Abazov:2005pn}. 
Several sub-detectors provide information to make a decision. 
At D\O\ the first level (L1) trigger system is a hardware system 
filtering the $2.5$ MHz beam crossing rate with a minimal dead time 
to an accept rate of about 2 kHz. A calorimeter trigger looks for energy 
depositions of high transverse energy; the central track trigger (CTT) and
the muon trigger provide tracks.

The second level trigger (L2) 
receives information from all major detectors to build 
a trigger decision using hardware and software algorithms. 
Each major D\O\ detector component
has a corresponding L2 preprocessor, the STT 
is the preprocessor 
of the Silicon Microstrip Detector. The information of all 
the L2 preprocessors are sent to the L2 global processor, which can
run filter algorithms to select the events and sends the information to
the third trigger level.
To maintain an acceptable dead time the 
mean decision time for L2 must be about $100\mu s$.  
The L2 output rate is about $1$ kHz. 
Finally, the third 
level software trigger partially 
reconstructs the events using a farm of processors and reduces the rate
to $50$ Hz, which is recorded for offline analysis.

\section{How does the STT work?}
The D\O\ tracking system consists 
of the Silicon Microstrip Tracker and the Central Fiber Tracker (CFT) 
both located in an about 2 Tesla solenoidal magnetic field.
The STT uses both devices to reconstruct the trajectories of charged particles.

In the first trigger level the Central Track Trigger  
can reconstruct tracks with a minimum $p_T$ of $1.5$~GeV
using information from the three scintillator-based detectors,
the Central Fiber Tracker and the central and forward preshower detectors.
The CFT consists of about 80000 scintillating fibers and 
because of its fast readout time its information can
already be used in the first trigger level.
The position resolution without the Silicon Tracker is, however, not
sufficient to provide precise information of the particle decay lengths. 

The SMT has six central barrels with four silicon
layers each and in total about 800000 readout channels.  
The barrel sensors parallel to the beam pipe are used for the STT. They
have a 50$\mu m$ pitch width. 
Offline tracks made using the SMT have an
impact parameter resolution of up to $15-20\mu m$ for high $p_T$ tracks,
 which is sufficient to detect B-mesons.

\begin{figure}
  \center
  \includegraphics[width=0.45\textwidth]{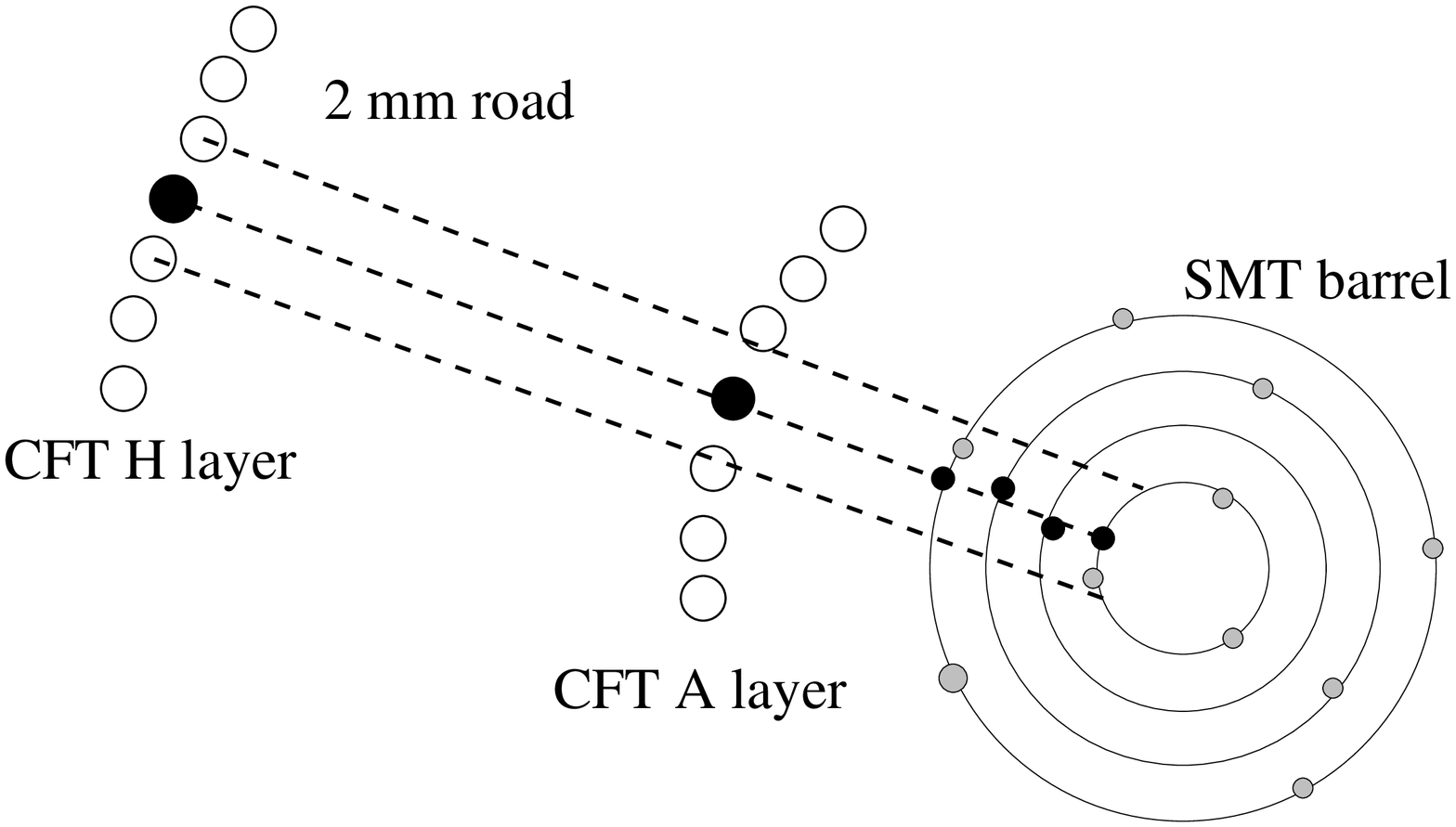}
  \caption{The 'road' as defined around the Level 1 track using the inner and outer layers hits of the CFT and the cluster selection in the SMT.
  }
  \label{fig:road}
\end{figure}

The STT uses the Level 1 tracks provided by 
the CTT as seeds to define 'roads' into the SMT (see Fig. \ref{fig:road}) .
These roads are cylinders of $\pm 2 \mbox{mm}$ radius around each CTT track. The STT
forms clusters from the pedestal subtracted 
SMT hits and only clusters within these roads are considered for track
fitting. Clusters are made by summing up a group of contiguous strips 
above some threshold. 

The STT design divides the SMT 
into 12 sectors, each 30 degrees in the azimuthal 
angle $\phi$ and the track fitting is
performed in parallel for each of the sectors. Almost no efficiency loss is
caused by tracks crossing sectors. The electronics for two sectors
house in one of the 6 STT crates. Data of the same sector of all 6 SMT
barrel detectors has to be routed to one crate.
Each crate has one Fiber Road Card which receives and distributes the Level 
1 tracks and communicates with the trigger framework. In each crate 9 Silicon Trigger Cards perform a pedestal correction, followed by the clustering of the SMT hits and associate the clusters to the roads. 

The information of the Fiber Road Card and Silicon Trigger cards is then
sent to the two Track Fit Cards. The Track Fit Cards receive the roads and
axial clusters and convert them via a large lookup table to physical coordinates. Then the two dimensional 
track fit is performed in the $r - \phi$ plane with the
form $\phi(r) = b/r + \kappa r + \phi_0$. Here $b$ is the impact parameter
with respect to the detector origin, 
$\kappa$ is the tracks curvature and $\phi_0$ is the
$\phi$ angle of the track at the distance of closest approach. 
The Track Fit Cards uses the clusters which are closest to the CTT track
and performs a fit if clusters are found in three or more SMT layers.
The track parameters are obtained by a integer $\chi^2$ method, which
minimizes the cluster-track residuals using 
matrices stored in an onboard lookup table. The track parameter is   
corrected for the beam spot position offset of the previous data 
taking run.
The output consists of the track parameters, the $\chi^2$ of the fit 
among other information.
The data is send to the L2 CTT preprocessor which merges it with Level 1
CTT information, formats the data and sends it to the L2 global processor.
The track parameters 
are provided on average 
in about $50\mu s$.  

\section{A B-identification algorithm for Level 2}
Combining tracks offline with B-identification algorithms greatly enhances
the performance of an experiment to detect B-mesons, obviously this may
also work at Level 2. A problem is that the current offline algorithms are not
fast enough to be run in less than $10-20 \mu s$.  
The algorithm described in the following  uses the STT tracks 
as input to an efficient multivariate $b$-identification and is 
fast enough to be run on the 
L2 global processor in less than $1-5 \mu s$. 

\begin{figure}
  \center
  \includegraphics[width=0.27\textheight]{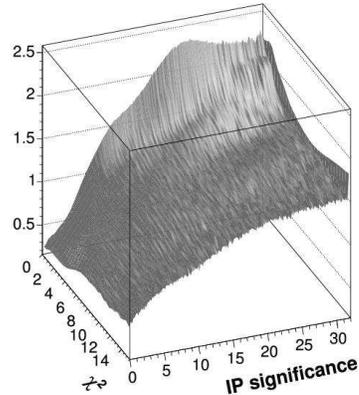}
  \caption{2-dimensional signal-to-background ratio as used in the B-identification algorithm.}
  \label{fig:MULM}
\end{figure}

The method uses the ratio of the probability density functions of the
signal prediction to the background prediction.
The signal is given by data events with a soft offline B-identification
, while background events are events without any offline B-identification.
The two-dimensional signal, $\mbox{pdf}_{i,S}$, and 
background, $\mbox{pdf}_{i,B}$, probability densities are derived 
with a smoothing method as a function of the track
impact parameter significance and the track $\chi^2$ of the track fit.
The impact parameter significance is given 
by $\mbox{IPsig}=IP/\sigma(IP)$, where $\sigma(IP)$ only 
includes the $P_T$ dependent effects of multiple Coulomb scattering and
$IP$ is the impact parameter.
Using $\chi^2$ in the $\mbox{pdf}$
downgrades badly fitted tracks without cutting 
them and loosing efficiency. Tracks with a larger 
$\chi^2$ still have limited lifetime information 
as can be seen from Figure~\ref{fig:MULM}.
For the trigger the signal-to-background ratios
$r(\mbox{IPsig},\chi^2)=\mbox{pdf}_{S}(\mbox{IPsig},\chi^2)/\mbox{pdf}_{B}(\mbox{IPsig},\chi^2)$ are stored in a  
lookup table.

The final discriminant is then derived by a loop over the 5 tracks
with highest impact parameter, 
accessing $r(\mbox{IPsig},\chi^2)$ for each track and
multiplying these values: 
\begin{displaymath}
D= \prod_{i=1}^5 r_i (\mbox{IPsig},\chi^2)
\end{displaymath}
 The event selection in done by cuts on this discriminant.

\section{Performance studies}
\begin{figure}
  \center
  \includegraphics[width=0.32\textheight]{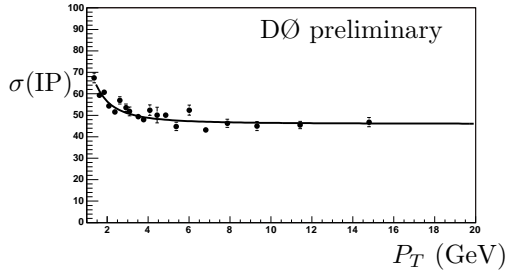}
\put(-100.0,80.0){D\O\ preliminary}
\put(-50.0,-5.0){$P_T$ (GeV)}
\put(-195,60.0){$\sigma$(IP)}
  \caption{STT impact parameter resolution (in $\mu m$) for a recent data run as a function of the $P_T$ of the track.}
  \label{fig:IPSIG}
\end{figure}

The STT is fully operational since May 2004. 
Figure \ref{fig:IPSIG} 
shows the impact parameter resolution of the STT as a function
of the $P_T$ of the track as determined with a recent data run. 
At the time of writing this article (Aug. 2005) a cut
on two tracks with an impact parameter 
significance $> 2$ or $>3$ and $\chi^2<5.5$ are implemented as 
L2 trigger requirement.

\begin{figure}
  \center
  \includegraphics[width=0.34\textheight]{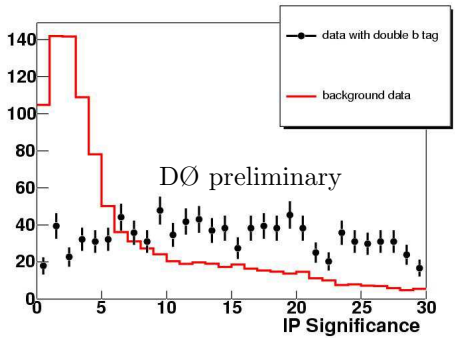}
\put(-120.0,60.0){D\O\ preliminary}
  \caption{Impact parameter significance distribution of the track with the largest impact parameter for signal and background data.}
  \label{fig:SIG2vtx}
\end{figure}

We show the STT B-identification performance by comparing
a dijet sample without any
offline b-tag (as background events) and a dijet sample 
where two secondary vertex tags and a muon tag are required (as signal events).
The impact parameter significance distribution for both samples using 
STT tracks with $\chi^2<5.5$
is shown in Fig \ref{fig:SIG2vtx}.

\begin{figure}
  \center
  \includegraphics[width=0.28\textheight]{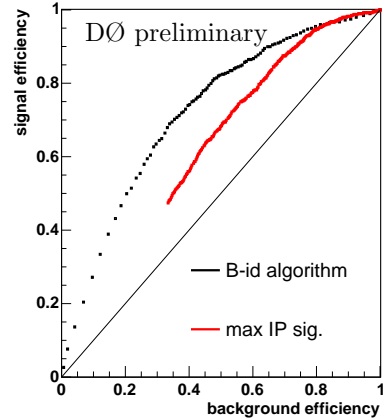}
\put(-128.0,155.0){D\O\ preliminary}
  \caption{Signal efficiency versus background efficiency for a cut on the maximum IP significance and on the discriminant of the B-id algorithm.}
  \label{fig:MULMonetag}
\end{figure}

Figure \ref{fig:MULMonetag} compares the signal and background efficiency 
for cuts 
on the IP significance and on the discriminant 
of the B-identification algorithm and for using 
STT tracks with $\chi^2<15.5$. Two secondary vertex tags are required for the
signal sample. 
The additional use of such an algorithm increases the background reduction
by up to a factor of 2 for the same signal efficiency. 
These findings are corroborated by using other offline B-selections as signal
events.

\section{Summary and Outlook}
The Silicon Track Trigger is a new component of the D\O\ second level trigger. It is since 2004 in smooth and stable operation
and allows to identify B-meson production by selecting 
events with a large impact parameter.
It has a high potential of enhancing the potential of 
several physics analyzes that rely on B-identification.
The implementation of a new SMT layer (Layer0) in the next shutdown period
will further increase the precision and stability of the STT.

\end{document}